\documentclass[manuscript]{aastex}

\slugcomment{}

\shorttitle{}
\shortauthors{Li et al.}

\begin{document}

\title{The Cross Spectral Time lags evolution along branches in XTE J1701-462}

\author{Zhao-sheng, Li\altaffilmark{1}}
\email{lizhaosheng@mail.bnu.edu.cn}

\author{Li, Chen\altaffilmark{1}}
\email{chenli@bnu.edu.cn}

\author{Jin-lu, Qu\altaffilmark{2}}
\email{qujl@ihep.ac.cn}

\author{Qing-cui, Bu\altaffilmark{1}}
\and
\author{De-hua, Wang\altaffilmark{1}}

\altaffiltext{1}{Department of Astronomy, Beijing Normal University, Beijing 100875, China}
\altaffiltext{2}{High-Energy Astrophysics Laboratory, Institute of High-Energy Physics, Chinese Academy of Sciences, Beijing 100039, China}

\begin{abstract}
We investigate the cross spectrum of XTE J1701-462 in various types of Neutron Star-Low Mass X-ray Binary subclasses, during its 2006-2007 outburst. We analyze the relation between the time lags and temporal variabilities. We find that the hard time lags accompany with the horizontal branch oscillations (HBOs) and the soft time lags dominate the noise in the low frequency range 0.1-10 Hz on HB. In the Cyg-like phase, the time lags decrease on the middle normal branch (NB) from HB/NB vertex to NB/FB vertex, whereas the time lags are roughly invariant in the Sco-like source. We discuss that the Compton upscattering by the corona introduces the soft lag in low frequency noise. We interpret that the variation of the Comptonization component from the disk emission lead to the HBOs' time lags evolution along the Z tracks. We also report the rms amplitude spectrum and phase lag spectrum for the NBO. The $\sim$ $160^{\circ}$ phase lag is found.  We present that the rms amplitude of both the Cyg-like and the Sco-like NBO linearly increase with the photon energy in low energy bands, and it will drop in the highest energy band.
\end{abstract}

\keywords{binaries: general --- stars : individual (XTE J1701-462) --- stars: neutron - X-rays: binaries --- X-rays: individual (XTE J1701-462) --- X-rays: stars }

\section{Introduction}
The Low mass X-ray binaries (LMXBs) are binary systems which contain a Neutron Star accreting  material
from a low mass companion star (see \citet{van06} for review). Based on the
correlated X-ray spectral and timing properties, they can be divided into two
subclasses, the Z sources with near Eddington luminosity ($\sim0.5-1~L_{\rm{edd}}$)
and the atoll sources with much lower luminosity ($\sim0.0001-0.5~L_{\rm{edd}}$).
The Color-Color Diagram (CD) and Hardness-Intensity Diagram (HID) of  Z source
have three branches, from top to bottom, which are so-called the horizontal branch (HB),
the normal branch (NB) and the flare Branch (FB) \citep{hasi89}, respectively. However, the CDs/HIDs of atoll sources are obviously different from those of the Z sources. A CD/HID of atoll sources also have three branches,  the extreme island state, the island state, and the banana state. Usually, a complete track of the Z sources
last from hours to days, whereas atoll sources last from  weeks to months.

In terms of the orientation of branches, the Z sources are classified
into the Cyg-like Z sources (\object{Cyg X-2, GX 5-1}, and \object{GX 340+0})
with a horizontal HB (``Z"-shaped tracks) and the Sco-like Z sources
(\object{Sco X-1}, \object{GX 17+2}, and \object{GX 349+2}) with a vertical
HB (``$\nu$"-shaped tracks). The Fast Fourier Transform (FFT) algorithm	was
applied in order to obtain the power density spectra (PDS) of LMXBs. In some
Z sources (\object{Cyg X-2},\object{GX 5-1}, \object{GX 340+0}, \object{GX 17+2},
and so on), the HB Quasi-periodic oscillation (QPO) signals, which vary with
frequencies in range from 13 Hz to 55 Hz,  are observed in the power spectra.
Almost all Z sources have 5-7 Hz QPO on NB and few of them have $\sim~20$ Hz QPO
on FB. The radiation hydrodynamic model \citep{lamb89,fort89,mil92} was proposed
to explain the $\sim$ 6 Hz NBOs. In this model, the NBOs of Z sources arise from
the optical depth oscillation of in-flowing medium near the neutron star scatting
by the outgoing X-ray flux. This model also explained a minimum in the NBO rms
amplitude spectrum when the phase lag spectrum showed $180^{\circ}$
phase jump. In \object{Cyg X-2}, a $150^{\circ}$ phase jump was observed around 6
keV, which was defined as pivot energy \citep{mit89,ste00}.
Moreover, the same phase jump was observed around 3.5 keV in \object{GX 5-1}.
Both of these two Cyg-like sources displayed a minimum around pivot energy in the
NBO rms amplitude spectrum. However, the \object{Sco X-1} which belongs
to Sco-like Z sources did not emerge a phase jump on the NB. The near-$180^{\circ}$
phase jumps observed in \object{Cyg X-2} and \object{GX 5-1} but not in \object{Sco X-1}
are consist with the classification of Z sources.

Two kinds of time lags were observed in LMXBs. Firstly, \citet{lei08} applied the
cross correlation function to \object{Cyg X-2} during its low-intensity states,
and detected both anticorrelation and positive correlation with hectosecond soft
and hard time lags. The anticorrelation and time lags were also observed in HB and
upper NB of \object{GX 5-1} \citep{sri12} which implied a truncated accretion disk
in LMXBs. Secondly, \citet{van87} used the Fourier cross spectrum technique to study
the \object{Cyg X-2} and \object{GX 5-1}. They found tens of milliseconds hard time
lags, i.e., the hard photons lag the soft photons in HBOs and NBOs, and soft lags
in low-frequency noise. \citet{kaa99} discovered millisecond soft lags in the atoll
source \object{4U 1636-536}. \citet{qu01} found that \object{Cir X-1} presents a hard
lag on the HB and a soft lag on the NB and FB. In order to explain both soft lags and
hard lags in LMXBs, several mechanisms, including a shot model \citep{vau94}, a uniform Comptonization model \citep{cui97}, a two-layer corona model \citep{nob01,qu01} and a disk propagation model \citep{utt11}, had been proposed.

The \object{XTE J1701-462} is a unique LMXB source observed by Rossi X-ray Timing
Explorer (RXTE) during its outburst in 2006-2007 \citep{rem06}. At high luminosity,
the \object{XTE J1701-462} switched between a Cyg-like Z source and a Sco-like Z
source and at low luminosity the transition from Z source to atoll source was
observed \citep{hom07,lin09,hom10}.
During these states, the shape of CDs/HIDs, the X-ray timing behavior
and the X-ray spectra also changed. It provides us a great opportunity to
understand the NBO behaviour, the time lag spectrum and the rms amplitude
spectrum, between Cyg-like phase and Sco-like phase. We also utilize the
cross spectrum to study the time lags evolution along with branches from Z source to
atoll source.

In Section 2, we describe the observations and analysis techniques and
our results are reported in Section 3. In Section 4 and 5, we present the discussion
and conclusions.

\section{Observations}

We analysed all 866 observations of the LMXB \object{XTE J1701-462} from
its 2006-2007 outburst collected with the Proportional Counter Array (PCA)
instrument on board of RXTE. During the outburst, the source showed X-ray
spectral and timing behaviours of the Z source and the atoll source.

\begin{figure}
\epsscale{.80}
\plotone{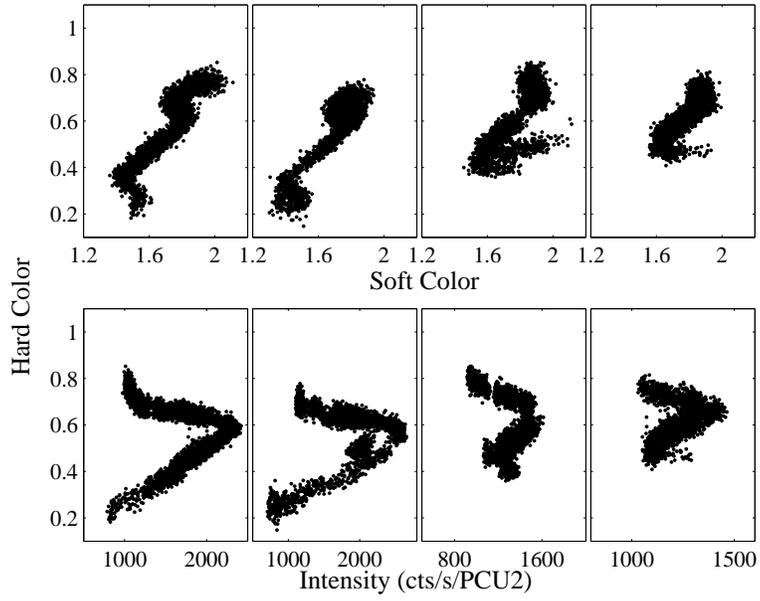}
\caption{CDs (upper panels) and HIDs (lower panels) for each of 4 intervals in table~\ref{tbl-1}. Each dot represents 16 s of background-subtracted data from PCU 2. From left to right, we label Interval I-IV. \label{fig1}}
\end{figure}

For all observations, we employed the following standard data selection criteria, a source elevation of $>10^\circ$, a pointing offset of $<0.01^\circ$ and a SAA exclusion time of 30 minutes. We utilized the Heasoft 6.12 to extract background-subtracted light
curves with a 16 s time resolution from the ``Standard 2" mode data and
no dead-time correction was applied. Only data from PCU 2 were used.
In order to obtain the CDs/HIDs from the light curves, we define the soft
color as 4.5-7.4 keV/2.9-4.1 keV (channels 10-17/6-9) count rates ratio, while the hard color as 10.2-18.1 keV/7.8-9.8 keV (channels 24-43/18-23) count rates ratio, then defined the intensity as count rates covering the energy range 2.9-18.1 keV (channels 6-43). Our data selection is listed in Table~\ref{tbl-1} and the CDs/HIDs
are displayed in Fig.~\ref{fig1}. The interval I and II belong to the
Cyg-like source, Interval III and IV belong to the Sco-like source.

We used the single-bit and science event model data with 32 s segments and $2^{-9}$ s time bin to created power density spectra (PDS) in frequency range 1/32 Hz-128 Hz. We studied the HBOs in energy range $\sim 2 - 60$ keV and the NBOs in energy range $\sim 7-60$ keV, because the NBOs were more apparent in $ \sim 7 - 60 $ keV than in $\sim 2 - 60$ keV \citep{hom07}. No background-subtracted and no dead time correction were applied. We employed the power spectral normalization of \citet{miy91}. And then, we fitted the power spectral using a multi-Lorentzian function \citep{now00,bel02} for the Interval I-IV. All parameters were set free \citep{pot03,str05}. We obtained the centroid frequency $\nu_{c}$, the full width half-maximum ($\Delta\nu$, FWHM) and the fractional root-mean-squared (rms) of HBO. We plotted the power spectra in the power times frequency representation ($\nu P_{\nu}$, in units of $rms^2$), where each power spectral density $P_{\nu}$ is multiplied by its Fourier frequency $\nu$.

We also extracted five energy bands, i.e., 2.06 - 4.49 keV (channels 0-10, \textbf{CH1}),
4.9 - 7 keV (channels 11-16, \textbf{CH2}), 7.4 -9.4 keV (channels 17-22, \textbf{CH3}),
9.8 - 14.8 keV (channels 23-35, \textbf{CH4}), 15 - 65 keV (channels 36-149, \textbf{CH5}), to compute cross spectra. The cross spectrum \citep{now99} is defined as $C(j)=S(j)^{*}H(j)$, where $S(j)$ and $H(j)$ are the measured complex Fourier coefficients at a given frequency $\nu_j$ for soft and hard energy bands, respectively. The phase lag between two energy bands light curves is the Fourier phase $\phi(j)={\rm{arg}}{[C(j)]}$. The time lag is constructed from $\phi(j)$ by dividing by $2\pi \nu_j$, i.e., $\tau(\nu_j)=\phi(j)/2\pi \nu_j$. A positive time lag indicates that the hard photons lag the soft photons. The cross spectra of \object{XTE J1701-462} show that the phase lag above 100 Hz are consistent with $\pi$. This is dead time effect which
should be corrected by subtracting a cross vector, averaged over 100 to 128 Hz \citep{van87,vau99}.
Hectohertz QPOs and kHz QPOs were observed in \object{XTE J1701-462} \citep{hom10,san10}, the
subtracted cross vector might be combined with the white noise and signal. However,
the signal above 100 Hz can be dismissed compared with the noise amplitude.
We computed the time lags of \textbf{CH2} - \textbf{CH5} relative
to \textbf{CH1} of NBO and investigated the time lags evolution along with HB
between 9.8 - 14.8 keV and 2 - 4.5 keV.

\begin{figure}
\epsscale{0.7}
\plotone{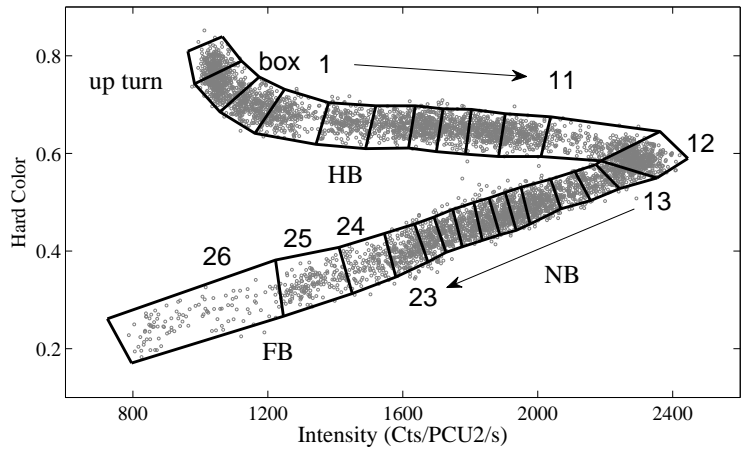}
\caption{The HID of Interval I. We divide it into 26 boxes from up turn to FB. We set the rank number $S_z$ of boxes 12 and 24 to 1 and 2, respectively. The $S_z$ of other boxes are fitted by spline interpolation. Each box contains $\geq 1500 ~\rm{s}$ of data.  \label{fig2}}
\end{figure}

To investigate the correlation between the data position in the HID and the temporal properties, we introduce a rank number $S_z$ defined by \citet{has90}. In Fig.~\ref{fig2}, we set the $S_z$ of box 12 and 24 as 1 and 2, respectively. The rank numbers of other boxes are decided by spline interpolation \citep{die00,lin12}.

\section{Results}

\subsection{Time lags evolution along HB in the Cyg-like phase}

\begin{figure}
\epsscale{.8}
\plotone{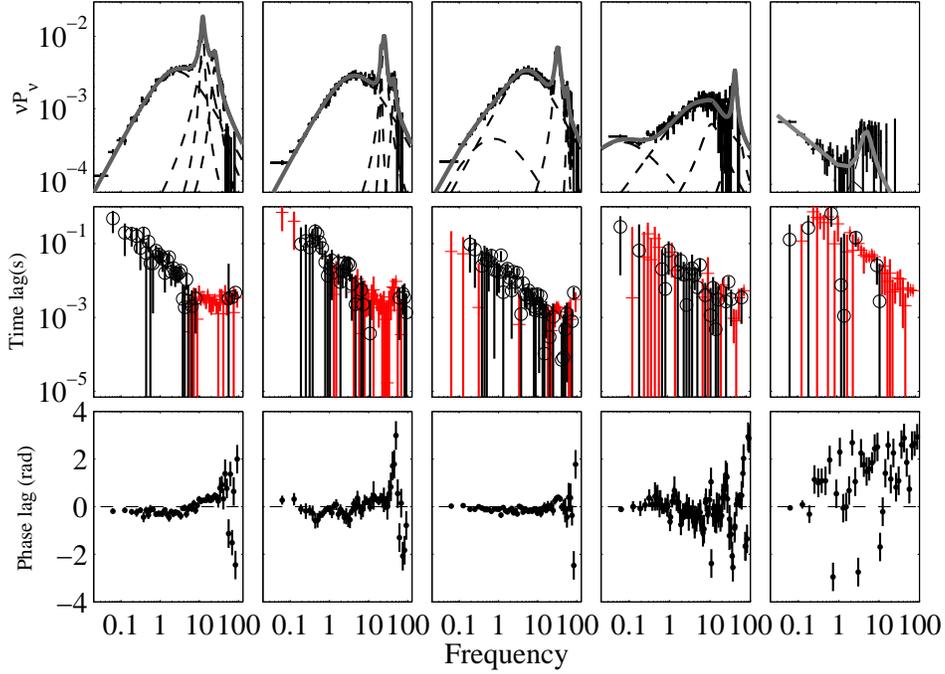}
\caption{Power density spectra (top panels, in units of $rms^2$) with 2-60 keV and the cross spectra (middle panels) between 9.8-14.8 keV and 2-4.5 keV on the up turn (box 1), the left HB (box 3), the middle HB (box 5), the HB/NB vertex (box 12) and the middle NB (box 18) from left to right of Interval I. The black circles and the red plus signs represent the soft and hard lag, respectively. The corresponding phase lags display in the bottom panels. The power density spectra are fitted by a multi-Lorentzian function. See text for the details. Each component presents as a dash line and the best fitting results are displayed as grey solid lines in top panels, which also appears in Fig.~\ref{fig4}-Fig.~\ref{fig5}.  \label{fig3}}

\end{figure}

\begin{figure}
\epsscale{.8}
\plotone{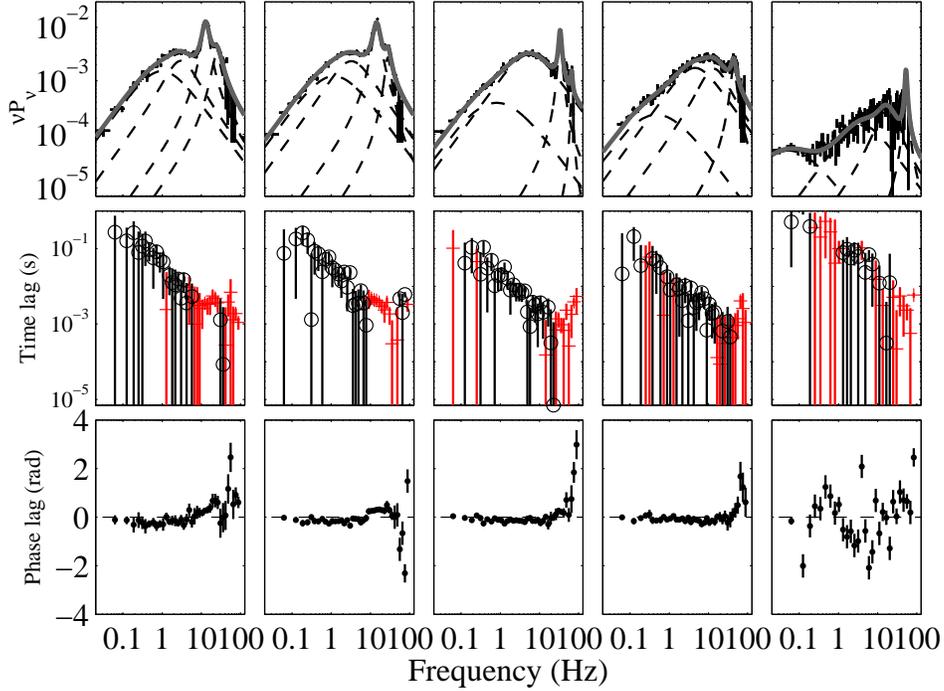}
\caption{Power density spectra (top panels) and the cross spectra (middle panels) as well as the phase lags (bottom panels) on the up turn, the left HB, the middle HB, the right HB and the HB/NB vertex  of Interval II. The selected energy channels are the same with Fig. ~\ref{fig3}. The PDSs and the cross spectra of the NB and FB are not provided here because the data are not dense enough. \label{fig4}}
\end{figure}

\begin{figure}
\epsscale{.8}
\plotone{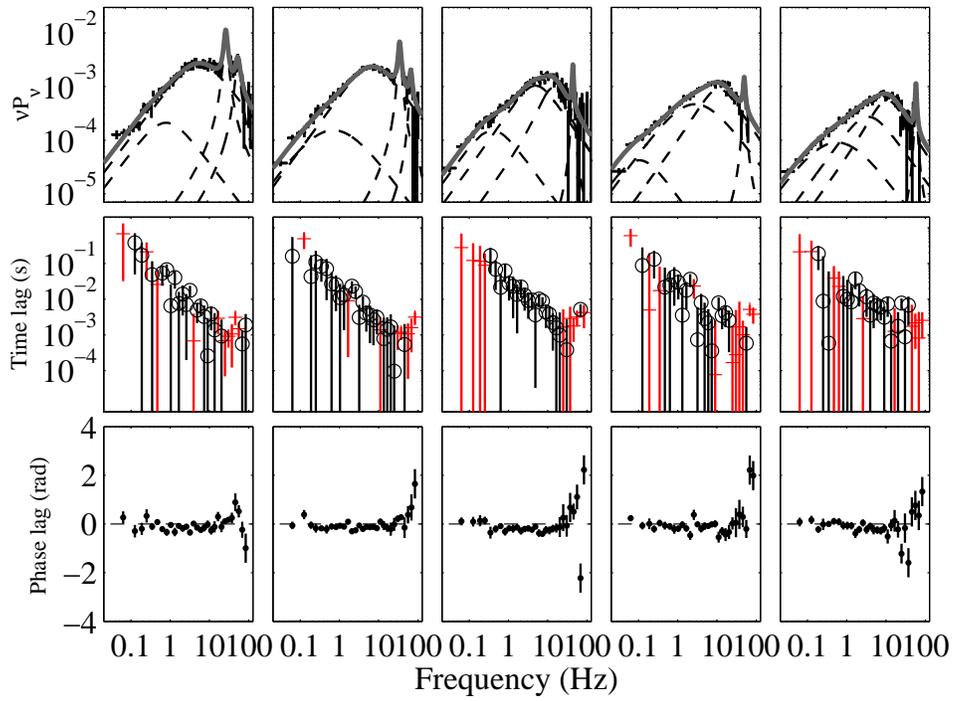}
\caption{The power density spectra (top panels), the cross spectra (middle panels) and the phase lags (bottom panels) on the HB of Interval III.\label{fig5}}
\end{figure}

We calculate the PDSs and average cross spectra for \object{XTE J1701-462} in the Cyg-like Z phase (Fig.~\ref{fig3} and Fig.~\ref{fig4}) and in the Sco-like Z phase (Fig.~\ref{fig5}). Positive lags indicate that the hard photons (9.8-14.8 keV) lag the soft photons (2-4.5 keV).
In Fig.~\ref{fig3}, the cross spectra of Interval I show hard lags (red `+') during QPOs' frequency range on HB. The red noises below 10 Hz display soft time lags (black `o') on HB. The time lags of NB are dominated by hard lags.  In Fig.~\ref{fig4}, the HBOs and cross spectra of Interval II are present. The time lags on HB of Interval II and Interval I have similar behaviours. In Fig.~\ref{fig5}, the hard lags are accompanied by QPOs and the soft lags are observed less than 10 Hz on HB. Meanwhile, we plot the phase lag versus frequency in Fig.~\ref{fig3}-\ref{fig5}, which clearly show the intrinsic phase lag couple with QPO appearing.

\begin{figure}
\epsscale{.80}
\plotone{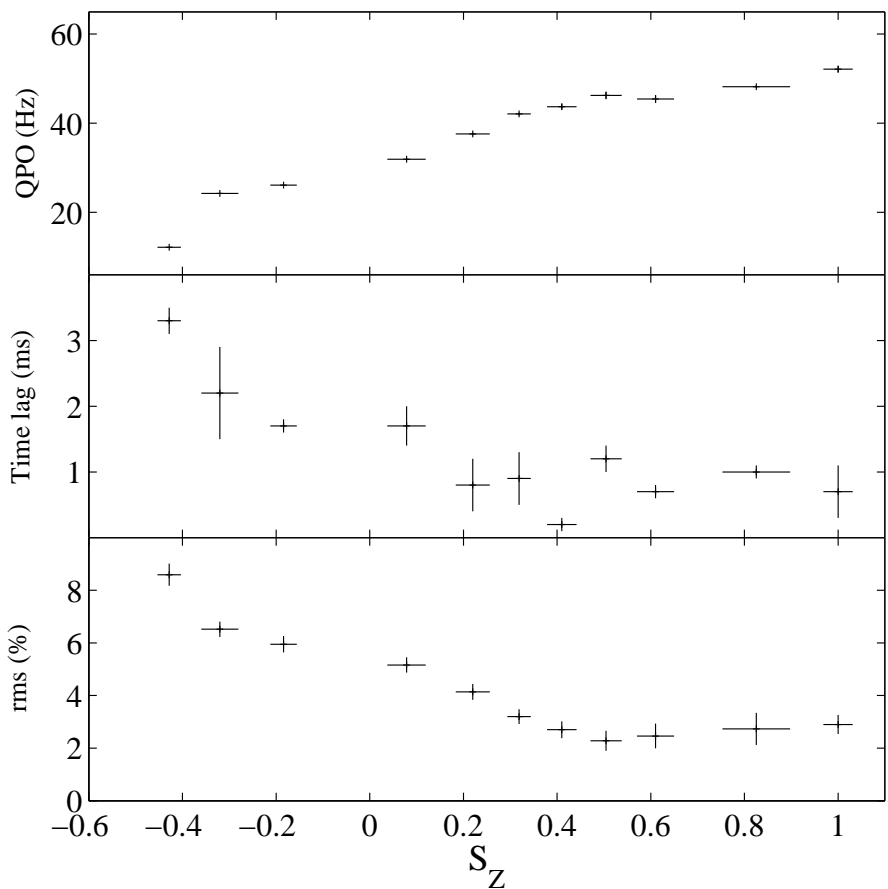}
\caption{QPO centroid frequency, the time lag of QPO and rms of QPO on the HB, as a function of $S_z$, of Interval I, from top to bottom. Positive lag means that hard photons lag soft photons.  The time lag of QPO is averaged between $\nu_{c}-\Delta{\nu}/2$ and $\nu_{c}+\Delta{\nu}/2$, where $\nu_{c}$ and $\Delta\nu$ are the centroid frequency and FWHM of QPO, respectively. \label{fig6}}
\end{figure}

\begin{figure}
\epsscale{.80}
\plotone{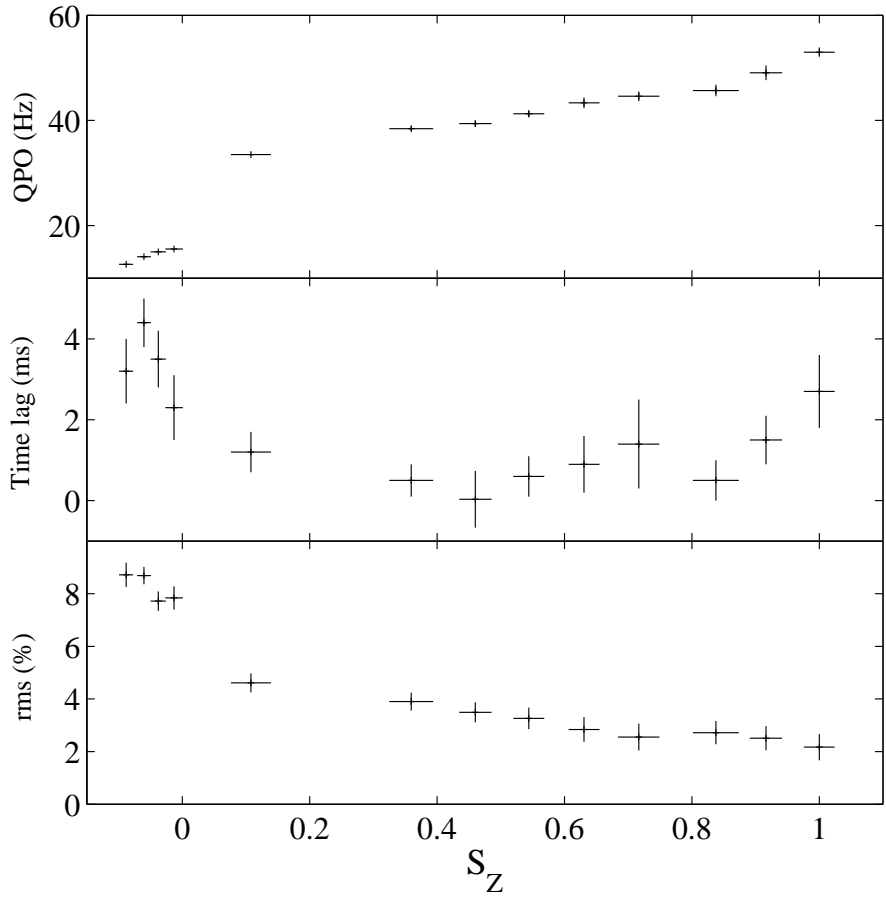}
\caption{QPO centroid frequency (top), the time lag of QPO (middle) and rms of QPO (bottom) on the HB of Interval II, as a function of $S_z$.   \label{fig7}}
\end{figure}

In order to investigate the correlation between $S_z$ and temporal variabilities, we average the time lags of QPOs over the frequency between $\nu_{c}-\Delta{\nu}/2$ and $\nu_{c}+\Delta{\nu}/2$.  In Fig.~\ref{fig6}, we display the QPO centroid frequencies, the time lags and rms of the HBOs as a function of $S_z$ in Interval I. The centroid frequencies of HBO increase from $12$ Hz to $53$ Hz, but the time lags and rms of HBO decrease when $S_z$ is below 0.5, then remain constant when $S_z$ is larger than 0.5. As a contrast, we provide the same temporal variabilities  of  Interval II in Fig.~\ref{fig7}. We notice that the centroid frequencies and the rms of HBOs are similar with interval I. However, The time lags rapidly decrease when $S_z$ less than 0.5, then the time lags slowly increase with $S_z$. All time lags are hard lags with several milliseconds.

\begin{figure}
\epsscale{.80}
\plotone{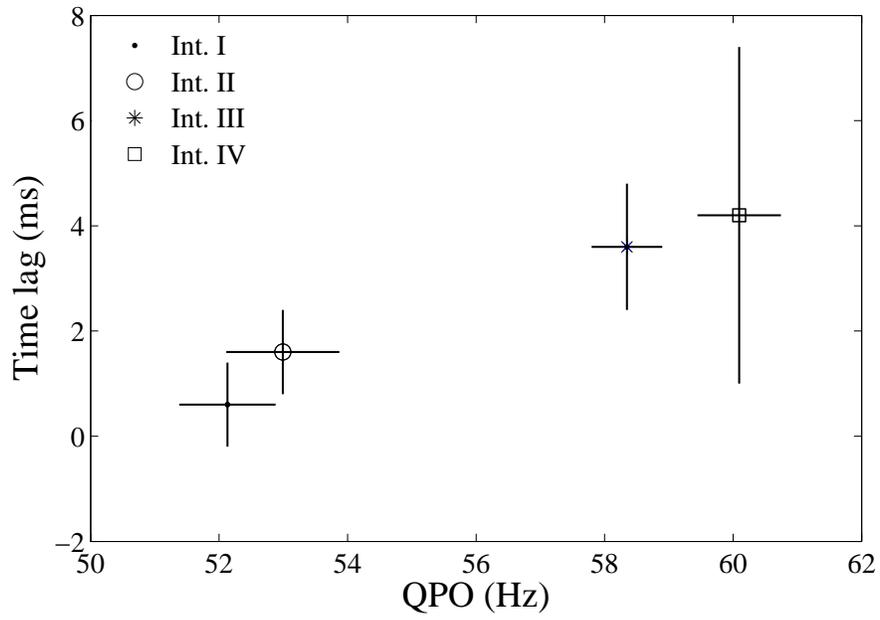}
\caption{The QPO centroid frequencies and the time lags of the HB/NB vertex for Interval I-IV from left to right, respectively. The HB/NB vertex region is manually selected from the relative HID. The time lag is computed between  7.4 - 9.4 keV and  2  -  4.5 keV.  \label{fig7-1}}
\end{figure}

For different HIDs, the intensity of the HB/NB vertex and the NB/FB vertex were considered as reference value for the accretion rate $\dot{m}$. In order to study the time lag - accretion rate connection,  we also calculate the time lag between 7.4 - 9.4 keV and  2  -  4.5 keV in the HB/NB vertex region of Interval I-IV. In Fig.~\ref{fig7-1}, the QPO centroid frequency positively correlates with the time lag. Both of them are  followed by source intensity decreasing, that is the accretion rate decreasing \citep{lin09}. We should notice that the source intensity difference between Interval I and Interval II are small (upper panel in Fig.~\ref{fig1}) and the time lags are extremely close too.

\subsection{Time lags evolution of the Z phase on NB}

\begin{figure}
\epsscale{1.0}
\plotone{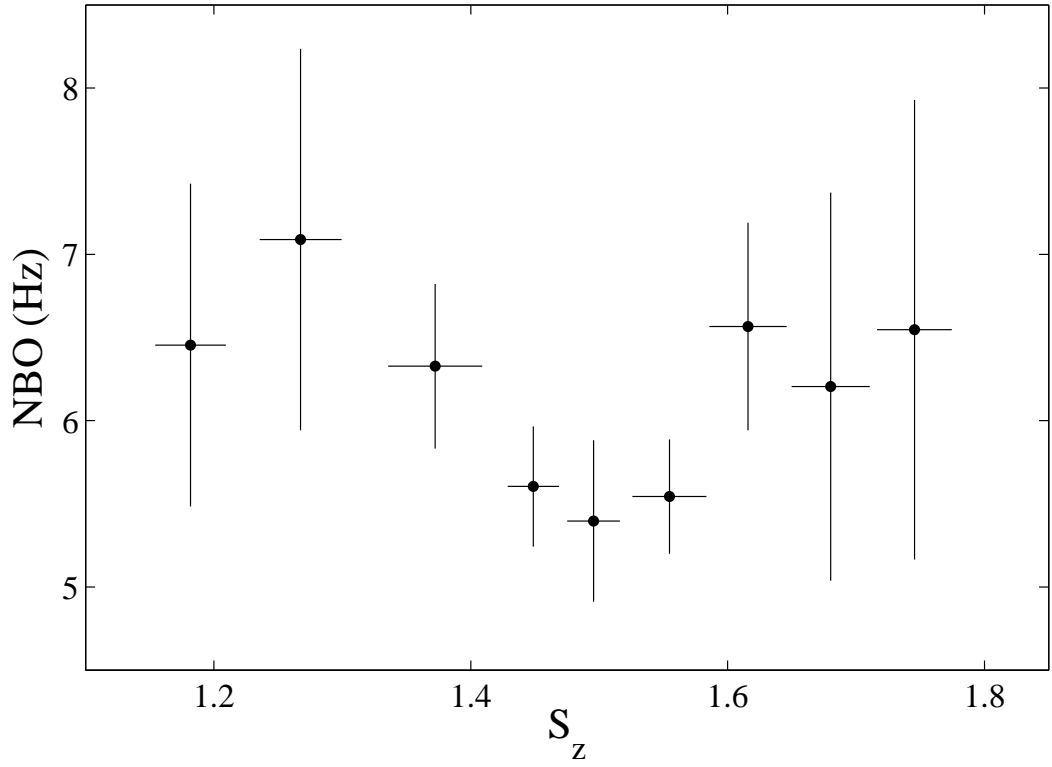}
\caption{The NBOs in Interval I as a function of $S_z$. The PDSs on the NB are computed from $\sim 7-60$ keV band light curves. The selection regions of NBOs correspond to the box 14-23 in Fig.~\ref{fig2}.   \label{fig7-2}}
\end{figure}

In Fig.~\ref{fig7-2}, we show the variation of NBO in Interval I. The centroid frequencies of NBO are around 6 Hz which consist with other Z sources.
In order to study the time lags along NB, we average the time lags with a fixed  frequency range 6$\pm$2 Hz, between 7.4 - 9.4 keV and 2 - 4.5 keV.

\begin{figure}
\epsscale{1.0}
\plotone{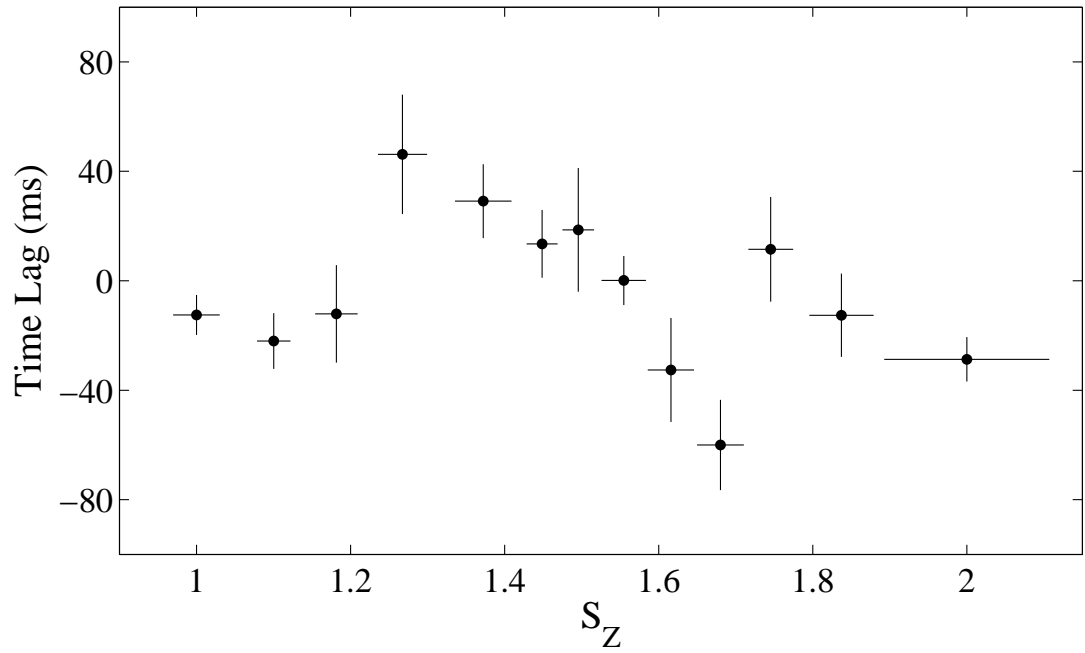}
\caption{Average time lags between 7.4-9.4 keV and 2-4.5 keV in the frequency between 4 and 8 Hz on the NB of Interval I, where the NBOs around 6 Hz were observed. \label{fig8}}
\end{figure}

In Fig.~\ref{fig8}, we display the time lags evolution from HB/NB vertex to NB/FB vertex of Interval I. The time lags are nearly constant on upper NB ($1\leq{S_z}<1.2$) and lower NB ($1.7\leq{S_z}\leq2$), and reduce from $\sim45$ ms to $\sim-70$ ms on middle NB ($1.2<S_z<1.7$).

\begin{figure}
\epsscale{1.0}
\plotone{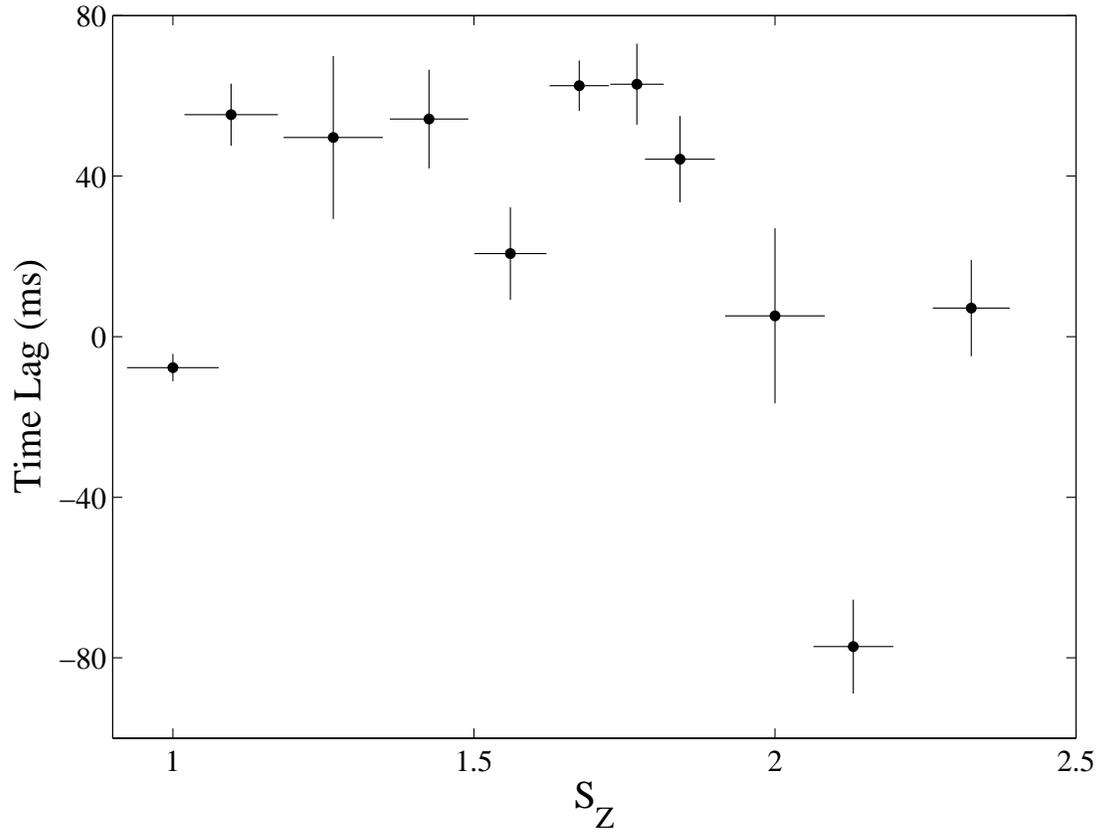}
\caption{Average time lags  on the NB and FB of Interval III. The energy channels and frequency range selection are the same with Fig.~\ref{fig8}. \label{fig9}}
\end{figure}

In Fig.~\ref{fig9}, the time lags as a function of $S_z$ on NB and FB of Interval III are reported. Between HB/NB vertex and NB/FB vertex, the time lags averaged over 4-8 Hz are nearly constant around 50 ms.

\subsection{ The time lags spectrum and rms amplitude spectrum of NBOs in the Cyg-like phase and the Sco-like phase}

We extracted the the Cyg-like phase and the Sco-like phase NBOs with five energy bands light curves,  2.06 - 4.49 keV (channels 0-10, \textbf{CH1}), 4.9 - 7 keV (channels 11-16, \textbf{CH2}), 7.4 -9.4 keV (channels 17-22, \textbf{CH3}), 9.8 - 14.8 keV (channels 23-35, \textbf{CH4}), 15 - 65 keV (channels 36-149, \textbf{CH5}), respectively. The NBOs' time lags of \textbf{CH2} - \textbf{CH5} against \textbf{CH1} are computed. For each energy band, the rms of NBO  is obtained by fitting the relative PDS with a fixed centroid frequency and FWHM. We defined the time lags as a function of energy  ``time lag (or phase lag) spectrum" and the rms as a function of energy  `` rms amplitude spectrum".

\begin{figure}
\epsscale{1.0}
\plotone{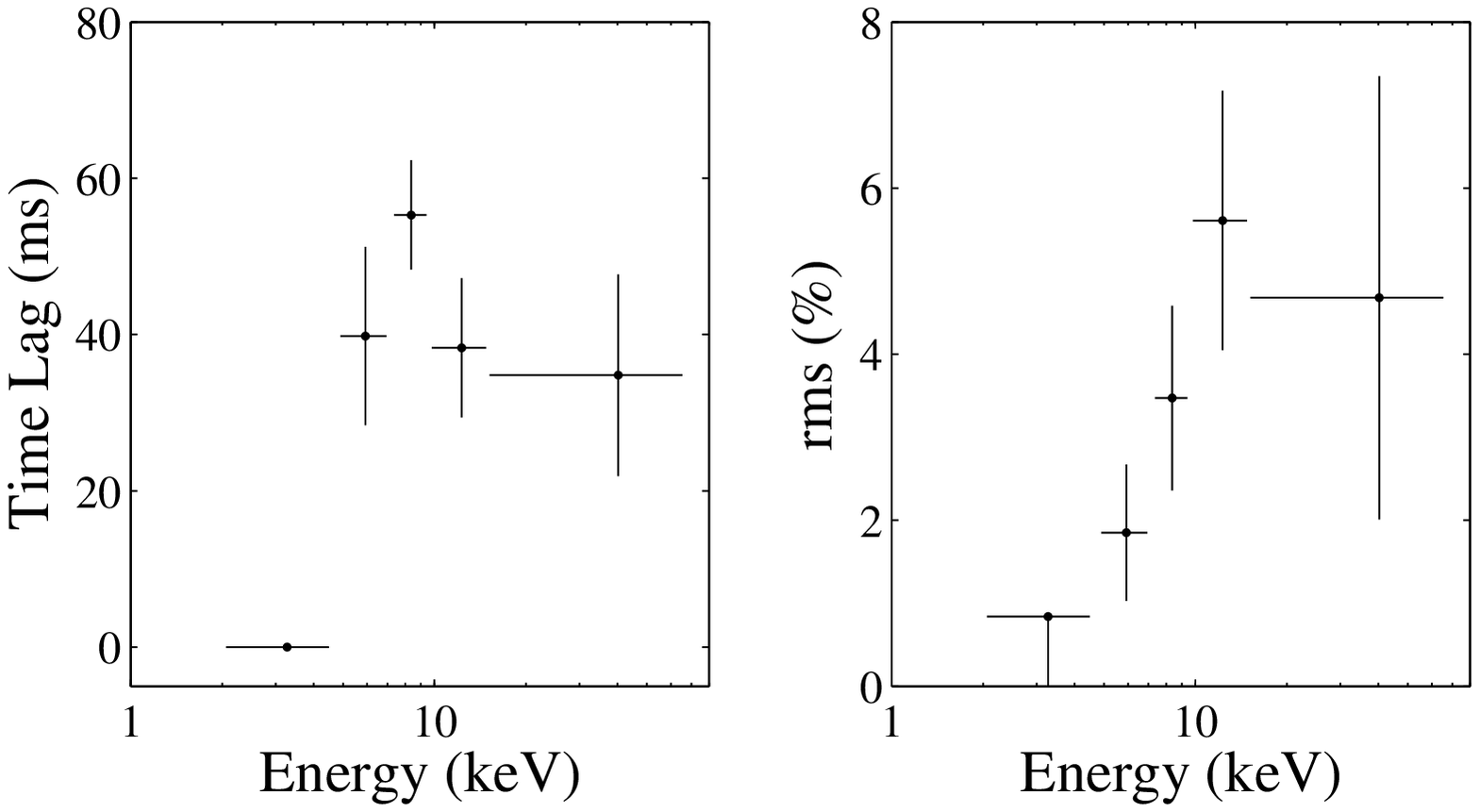}
\caption{The time lag spectrum (left panel) and the rms amplitude spectrum (right panel) for the NBOs on the lower NB of Interval I.\label{fig10}}
\end{figure}

\begin{figure}
\epsscale{1.0}
\plotone{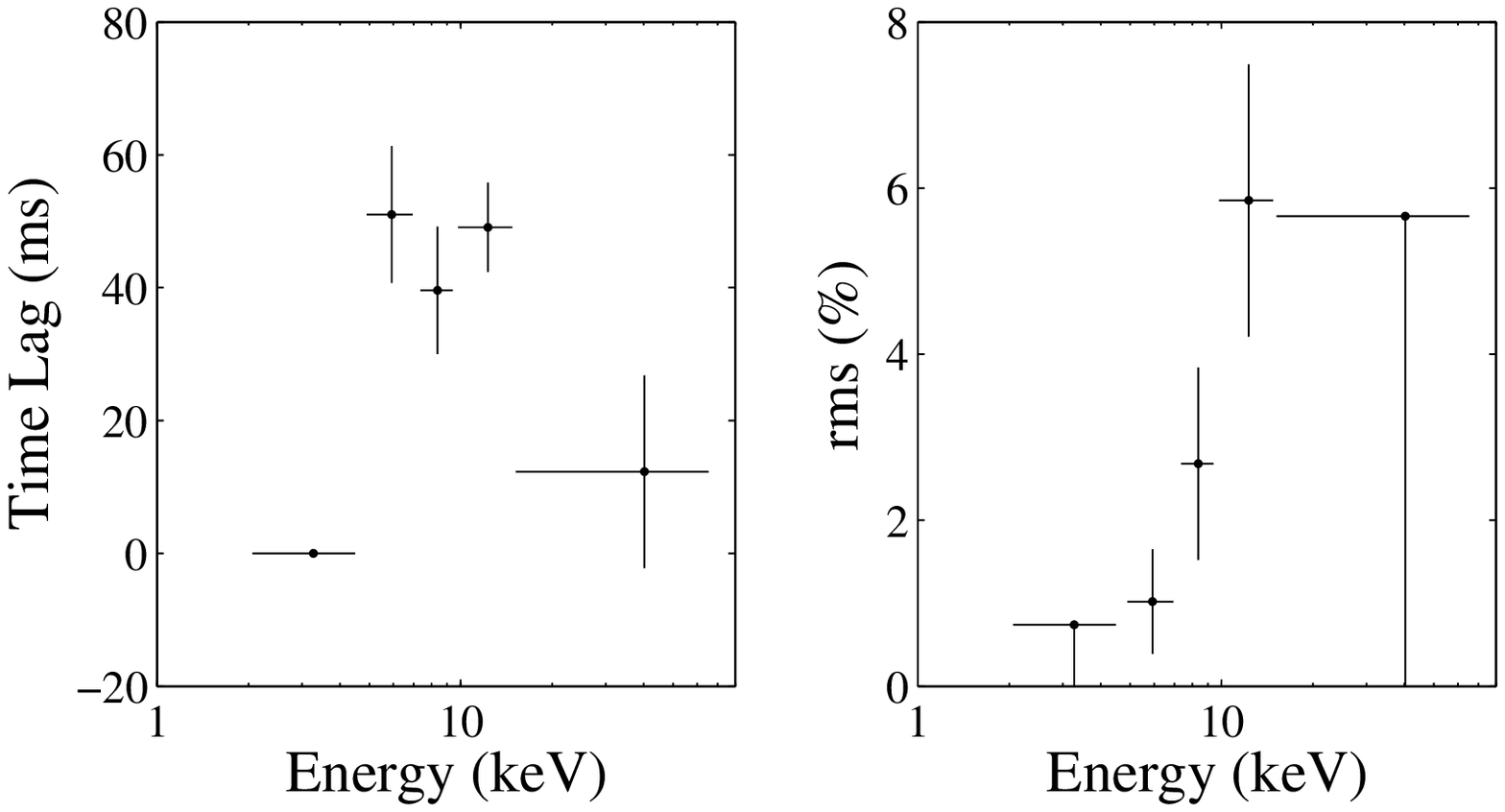}
\caption{The time lag spectrum (left panel) and the rms amplitude spectrum (right panel) for the NBOs on the lower NB of Interval III. \label{fig11}}
\end{figure}

\begin{figure}
\epsscale{1.0}
\plotone{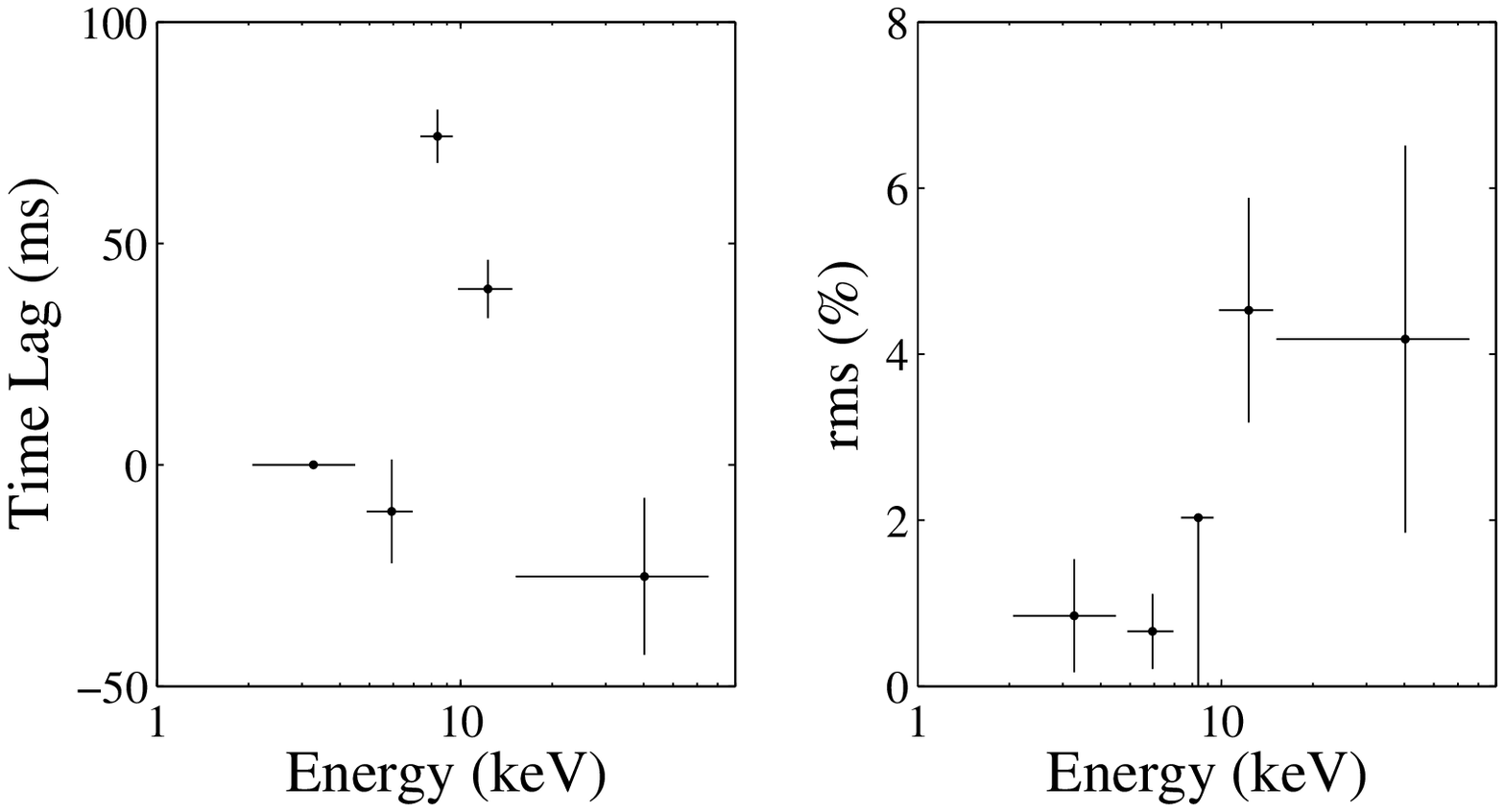}
\caption{The time lag spectrum (left panel) and the rms amplitude spectrum (right panel) for the NBO on the lower NB of Interval IV.\label{fig12}}
\end{figure}

In Fig.~\ref{fig10}, we notice that the maximum time/phase lag on the lower NBO is 55$\pm$7 ms or $\rm{114}\pm\rm{15}^{\circ}$ in Interval I, respectively. On the lower NB of Interval III , the time/phase lag is 49$\pm$6 ms or $\rm{106}\pm\rm{13}^{\circ}$ (Fig.~\ref{fig11}). On the lower NB of Interval IV, the time/phase lag are slightly larger than  Interval I and III, i.e.,  74$\pm$16 ms or $\rm{160}\pm\rm{13}^{\circ}$ (Fig.~\ref{fig12}). The near-$160^\circ$ phase jumps occur around 10 keV in both the Cyg-like phase and the Sco-like phase.

 Regardless of \textbf{CH5}, the rms amplitude of NBOs linearly correlate with energy in the range 2 - 15 keV (i.e., the signal strength of QPOs are proportional to the photon energy.) of Interval I and III (right panels in Fig.~\ref{fig10} and Fig.~\ref{fig11}). The rms energy dependence was also observed in \object{Sco X-1} \citep{wan12}. Because of low count rates and low signal-to-noise in the highest energy band, the rms of NBOs have large error bars. The drop of the rms amplitude in the highest band was appeared.  A pivot energy in 4.9-7 keV appears in the rms amplitude spectrum of Interval IV (right panel in Fig.~\ref{fig12}) but with a weak tendency.

\section{Discussion}

The LMXB source \object{XTE J1701-462} transited from a Z source into  an atoll source during its outburst. We used the FFT to compute the PDS of \object{XTE J1701-462}. For the Cyg-like observations (Interval I \& II), the centroid frequencies of HBO increase from 12 Hz to 53 Hz, while the amplitude of HBO becomes weaker. The cross spectra of HBO were computed as well. In Fig.~\ref{fig3} and Fig.~\ref{fig4}, we found that the time lags of low frequency noise less than 10 Hz are dominated by soft lags. Moreover, the soft lag increases towards low frequency. When QPOs appeared, the \object{XTE J1701-462} showed hard lags, i.e., high energy photons lag soft photons. For the NB observations of the Cyg-like phases and the Sco-like phases, the hard lags were also observed during NBOs' frequency range. These also appeared in other Z sources \citep{van87,vau94,qu01,qu04}.
We also study the time lags evolution along NB in the Z sources. In Interval I, the time lags averaged between 4 Hz and 8 Hz decrease on middle NB as a function of rank number $S_z$ (Fig.~\ref{fig8}). This could be the centroid frequency and rms amplitude variation of the NBO (Fig.~\ref{fig7-2}).  In interval III, the time lags of NBO nearly remained 50 ms (Fig.~\ref{fig9}).  For the HB/NB vertex regions in the Z sources, the time lag increased with the source intensity decreased, which is the trend of accretion rate decreased.

In order to explain the observed time lag in compact star system(e.g., BH-, NS-LMXBs, AGNs, and so on), many models are proposed, including the shot model \citep{vau94}, the propagation of accretion fluctuations model \citep{utt11}, and the Comptonization models \citep{cui97b,kaz97}. The shot model mathematically explained the observed power spectra in LMXBs \citep{alp85}, the soft lag in the low frequency noise \citep{shi87} and the hard lag in the HBO simultaneously \citep{vau94}. It regarded the HBO as the signature of intensity modulation and the low frequency noise as the power spectral signature of shots, when the accretion matter fall onto the Neutron Star surface. The observed soft lag of low frequency noise could be interpreted if the shot envelopes in the soft energy band lag those in the hard energy band. Meanwhile, the hard lag in the HBO may originate from a particular delaying of the QPO production in high energy band without distinctly interfering by low frequency noise.

The time lag in the hard state of BH-LMXBs which displays hard lag in low frequency noise shows the exactly opposite to our results \citep{now99,pot03,utt11}. \citet{now99} investigated the hard state time lag in the Cyg X-1 and found that the energy dependence of hard lag in low frequency noise as well as the hard lag increase with decreasing low Fourier frequency. It seems contradictory to the uniform and compact Comptonization corona model. \citet{utt11} pointed out that the observed hard lag in GX 339-4 corresponds to the viscous propagation of mass accretion instabilities through the disk. The emission of BH-LMXBs contains two parts, the power law dominant in the high energy band and the disk black body dominant in the low energy band. When the mass accretion fluctuations happened, the variabilities of disk black body emission lead to the power law emission variations on time scales of seconds. If the low frequency noise comes from these two components variations, then the hard lag appears. For the soft lag, it may be responsible by the power law illuminating the black body on time scales of milliseconds, where the the power law emission regions locate about tens of $R_G$ to the disk.

However, XTE J1701-462 shows more complicated emission mechanism than BH-LMXBs \citep{lin09,ding11}. On the HB, the energy spectral of XTE J1701-462 is composed of the multi-color black body from the disk, the black body from the NS surface and the cutoff power law from the corona above the disk, which dominate the emission in several keV, tens of keV and the hard tail respectively. On the NB and FB, the cutoff power law component becomes too faint to be detected. We indicate that there exist three possible lag mechanisms, the Compton upscattering by the corona \citep{kot01,are06}, the propagation of the accretion rate instabilities in the disk \citep{utt11}  and the disk  illuminating by the  black body from the NS surface in XTE J1701-462. However, the soft lag introduced by the NS X-ray illuminating is likely the time scale of light travel from the NS to the disk, which is too low ($\sim0.1~ms$) compared to our calculations. The propagation of the accretion rate instabilities in the disk only produce hard lag in low frequency noise which is not conform to the soft lag in XTE J1701-462.

The uniform Comptonization models \citep{pay80,kaz97} can not interpret the energy dependence of hard time lag and soft time lags in LMXBs.
In order to explain soft lags and time lags evolution of QPOs, a Comptonization
model with two-layer corona was proposed \citep{nob01,qu01}. When photons pass through the two-layer corona, the photons encounter the inverse Compton scattering by hot electrons and gain energy.  If the inner layer is hot and optical depth is $<1$, the hard photons are scattering more than the soft photons and the hard time lags observed. If the inner layer is optical thick, the photons are efficiently upscattered. When the harder photons pass through the out part corona with lower temperature, they will lose energy and the soft lags appear.

\citet{qu01} explained that the accretion rate changing along branches leads to the \object{Cir X-1} time lag evolution. From multiwavelength campaigns of Cyg X-2, the monotonically increasing in $\dot{m}$ from the HB, via the NB, to the FB was reported \citep{has90,vrt90}. If the accretion rate of XTE J1701-462 arise from the HB to the NB, the hot electrons in the inner layer corona are cooled by photons from the disk. This process will shrink the inner lay corona and the observed HBO time lags become smaller.
Lin and co-workers \citep{lin09,hom10,lin12} concluded that the accretion rate did not vary significantly along the Z tracks but did along secular change, i.e. $\dot{m}$ decrease from the Z source to an atoll source. The time lag - QPO relation of the HB/NB vertex could be explained by the accretion rate decreasing (Fig.~\ref{fig7-1}). However, for the time lags evolution on the HB of XTE J1701-462, when the source intensity increased, the disk luminosity which was mainly contributed by Comptonization ascended simultaneously (see Fig. 20 in \citet{lin09}).  When the luminosity of the disk component increases from the up turn to the HB/NB vertex, the temperature of the inner layer corona cool down through Compton emission and then lead to a smaller radius layer. Because the photons contributing to the HBO came from the disk, the time lags of HBO decrease as a function of $S_z$ was also observed. In a Z track, the time lag of HBO increasing towards high source intensity can be explained by the accretion rate increasing and the Comptonization component variations. \citet{jac09} and \citet{chu10} suggested an opposite accretion rate increasing direction, i.e., from the NB to the HB, which could not reasonably explain the time lags evolution of XTE J1701-462.

The near-180$^\circ$ phase lag and the dip in rms amplitude spectrum were observed in \object{Cyg X-2} and \object{GX 5-1}, but did not show in \object{Sco X-1}. \citet{ste00} provided two explanations for the NBOs behavior for \object{Cyg X-2} and  \object{GX 5-1} and \object{Sco X-1}. Firstly, they may have different viewing angle. Secondly, with the radiation-hydrodynamic model for the QPOs, the QPO of \object{GX 5-1} and \object{Sco X-1} are determined by optical depth oscillations while the QPOs of \object{GX 5-1} are determined by luminosity variation, where the NS of \object{GX 5-1} has a weaker magnetic field. \object{XTE J1701-462} changes from a Cyg-like Z source, via a Sco-like Z source, to an atoll source in two years. The binary inclination and the strength of the NS magnetic field are invariant on such a short timescale \citep{kuu94,psa95}. The accretion rate dominates the whole processes. We found near-160$^\circ$ phase lag of  NBOs  both in Cyg-like phase and Sco-like phase which are consistent with \object{Cyg X-2} and \object{GX 5-1}. However, no pivot energy in rms amplitude spectrum was observed in the Cyg-like phase. This can not  be explained by the radiation-hydrodynamic model \citep{fort89}. \citet{wan12} interpreted that the NBO and the rms energy dependence in Sco X-1 originate from the transition layer. The similar behaviors of time lag spectrum and rms amplitude spectrum between the Cyg-like sources and the Sco-like sources indicate that the NBOs are irrelevant to the luminosity variation in \object{XTE J1701-462}.

\section{Conclusions}
We investigated the cross sprctral and the QPOs for the Cyg-like phase and the Sco-like phase of XTE J1701-462. The QPOs increase from 12 Hz to 53 Hz on the HB and remain $\sim 6$ Hz on the NB. These consist with other Z sources. We also computed the time lags and HBOs for the HB/NB vertexes (Interval I-IV). The two-layer corona Comptonization model can explain the soft lag in low frequency noise and hard lag in HBO respectively. Both the accretion increasing and the Comptonization component variation can produce the HBOs' time lag evolution. For a NS-LMXB, we could not deduce its accretion rate increasing direction along branches in the HID only from the time lag variation. The spectral states should be analyzed simultaneously. This will be discussed in our future work.

\acknowledgments
We would like to appreciate the referee for valuable comments and suggestions. We are also grateful to M.Y., Ge, Y.N., Wang, H.Q., Gao and Z., Zhang for data reduction. Project supported by the National Natural Science Foundation of China (10778716); the National basic Research program of China 973 Program 2009CB824800; the National Natural Science Foundation of China (11173024); the Fundamental Research Funds for the Central Universities. This research has made use of data obtained from the High Energy Astrophysics Science Archive Research Center (HEASARC), provided by NASA's Goddard Space Flight Center.

\clearpage

\clearpage

\begin{deluxetable}{cccccc}
\tabletypesize{\scriptsize}
\tablecaption{The data selection of the CDs and HIDs.\label{tbl-1}}
\tablewidth{0pt}
\tablehead{
\colhead{Interval}    & \colhead{Begin of date } & \colhead{Begin of Obs.} &
\colhead{End of date} & \colhead{End of Obs.}   & \colhead{Source Type}   \\
\colhead{}            & \colhead{(DD/MM/YY)}              &  \colhead{}   &
\colhead{(DD/MM/YY)}            & \colhead{}    &  \colhead{}            }
\startdata
  I &   21/01/06 & 91106-01-04-00 &   31/01/06 & 91106-02-02-10 &   Cyg-like \\

 II &   02/02/06 & 91106-02-02-14 &   12/02/06 & 91442-01-01-02 &   Cyg-like \\

 III &  17/02/06 & 91442-01-07-02 &   25/02/06 & 91442-01-03-03 &   Sco-like \\

 IV &   08/03/06 & 92405-01-01-05 &   15/03/06 & 92405-01-02-05 &   Sco-like \\
\enddata
\end{deluxetable}

\clearpage

\clearpage

\clearpage

\clearpage

\clearpage

\clearpage

\clearpage

\clearpage

\clearpage

\clearpage

\clearpage

\clearpage

\clearpage

\clearpage

\end{document}